\pdfoutput=1
\def\year{2020}\relax
\documentclass[letterpaper]{article} %
\usepackage{aaai20}  %
\usepackage{times}  %
\usepackage{helvet} %
\usepackage{courier}  %
\usepackage[hyphens]{url}  %
\usepackage{graphicx} %
\usepackage{graphicx}  %
\usepackage{comment}
\usepackage{color}
\usepackage{xcolor}
\usepackage{bm,amsmath,bbm}
\usepackage{amsfonts}
\usepackage{amsthm}
\usepackage{subcaption}
\usepackage{multirow}
\usepackage{makecell}
\usepackage{enumitem}
\usepackage[normalem]{ulem}
\usepackage{varwidth}
\usepackage{array}
\DeclareMathOperator*{\minimize}{minimize}
\DeclareMathOperator*{\subjectto}{subject\, to}

\title{CrowDEA: Multi-view Idea Prioritization with Crowds}
\author{Yukino Baba\\
Unviersity of Tsukuba\\
baba@cs.tsukuba.ac.jp
\And Jiyi Li\\
University of Yamanashi\\
jyli@yamanashi.ac.jp
\And Hisashi Kashima\\
Kyoto University\\
kashima@i.kyoto-u.ac.jp
}
 
\begin{document}

\maketitle
\begin{abstract}
Given a set of ideas collected from crowds with regard to an open-ended question, how can we organize and prioritize them in order to determine the preferred ones based on preference comparisons by crowd evaluators?
As there are diverse latent criteria for the value of an idea, multiple ideas can be considered as ``the best''. 
In addition, evaluators can have different preference criteria, and their comparison results often disagree.
In this paper, we propose an analysis method for obtaining a subset of ideas, which we call frontier ideas, that are the best in terms of at least one latent evaluation criterion. 
We propose an approach, called CrowDEA, which estimates the embeddings of the ideas in the multiple-criteria preference space, the best viewpoint for each idea, and preference criterion for each evaluator, to obtain a set of frontier ideas.
Experimental results using real datasets containing numerous ideas or designs demonstrate that the proposed approach can effectively prioritize ideas from multiple viewpoints, thereby detecting frontier ideas. 
The embeddings of ideas learned by the proposed approach provide a visualization that facilitates observation of the frontier ideas. In addition, the proposed approach prioritizes ideas from a wider variety of viewpoints, whereas the baselines tend to use to the same viewpoints; it can also handle various viewpoints and prioritize ideas in situations where only a limited number of evaluators or labels are available. 
\end{abstract}

\section{Introduction}
Despite the recent advances in artificial intelligence, there are still several challenges that humans can handle better than machines, especially abstract, open-ended, and context-dependent problems.
Brainstorming new ideas is a typical example; for instance, to answer open-ended questions, such as ``What is the best logo for the next summer Olympic games?'', ``How can we reduce the number of latecomers at team meetings'', and ``What are the most reasonable solutions for preventing global warming?'', humans are expected to present more creative and reasonable solutions than machines.
Existing studies demonstrate that crowdsourcing is an effective approach to collecting several creative ideas from a wide range of people~\cite{yu2011cooks,koyama2014crowd,siangliulue2015toward,prpic2015work}.

Let us consider the example of designing a suitable logo for the next Olympic games. For example, let us assume that we ask crowd workers to provide a set of candidate designs. After collecting several design ideas, we should organize and prioritize them to select the best. However, the criteria for the best design are usually multi-faceted; for example, there may be two different criteria for design, e.g.,  traditional aesthetics and contemporary aesthetics. Therefore, there rarely exists a single overwhelming winner over the other candidates in terms of all criteria. Moreover, it is often difficult to define the criteria in advance. 

Thus, we must turn to the crowd for assistance, with the expectation that crowd evaluators may be able to identify the unknown diverse criteria. We must ask them to evaluate the ideas, often in the form of pairwise preference comparisons. 
The criteria for these comparisons can also be diverse depending on evaluators' personal viewpoints. 

In this study, we consider the problem of aggregating the pairwise idea preference comparisons by crowds containing different viewpoints so that a set of best ideas from certain viewpoints may be obtained. These ideas are called {\it frontier ideas}.
The proposed method, which is called \textsc{CrowDEA}, generates a priority map that is a low-dimensional latent space, where ideas are embedded such that the frontier ideas are furthest from the origin and the ideas projected onto the viewpoint of each evaluator are consistent with their pairwise comparisons.

Existing studies~\cite{bt,2dbt,crowdbt} estimate a unique rank list from the pairwise preference comparisons; they usually assume that there exists a unique rank list as the ground truth. 
In addition, as there are no explicit evaluation criteria readily available, existing methods, such as skyline query~\cite{borzsony2001skyline,hose2012survey,crowdskyline}, cannot be used.
The priority map of \textsc{CrowDEA} assists in making the final decision or further analysis (such as next-round idea sourcing) by providing an organized view from various perspectives. 

\begin{figure*}
  \begin{minipage}[b]{0.55\textwidth}
    \centering
    \subcaptionbox{Target objects (ideas): balls with different sizes and colors.\label{fig:objects}}{
    \includegraphics[height=0.8cm]{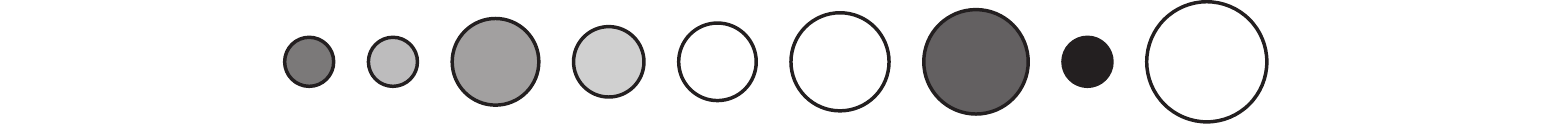}}
    \vfill
    \centering
    \subcaptionbox{Input: pairwise comparison results\label{fig:pairwise}}{
    \includegraphics[height=1.6cm]{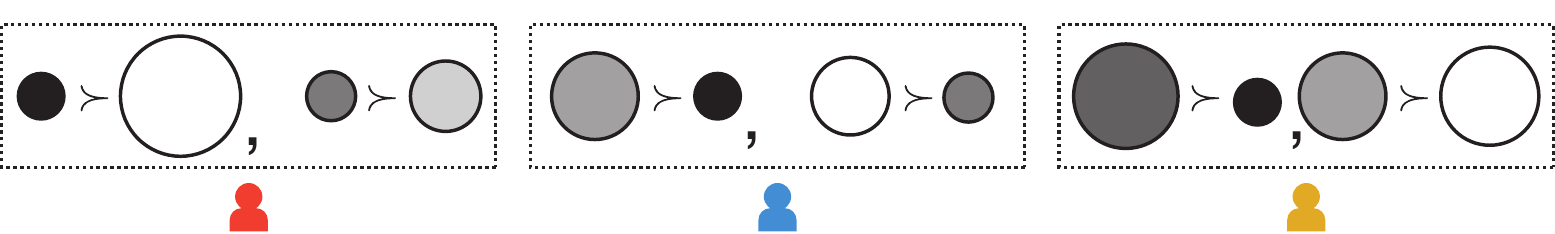}}
    \end{minipage}
    \hfill
    \begin{minipage}[b]{0.4\textwidth}
        \centering\subcaptionbox{Output: priority map\label{fig:priority_map}}{\includegraphics[height=3.8cm]{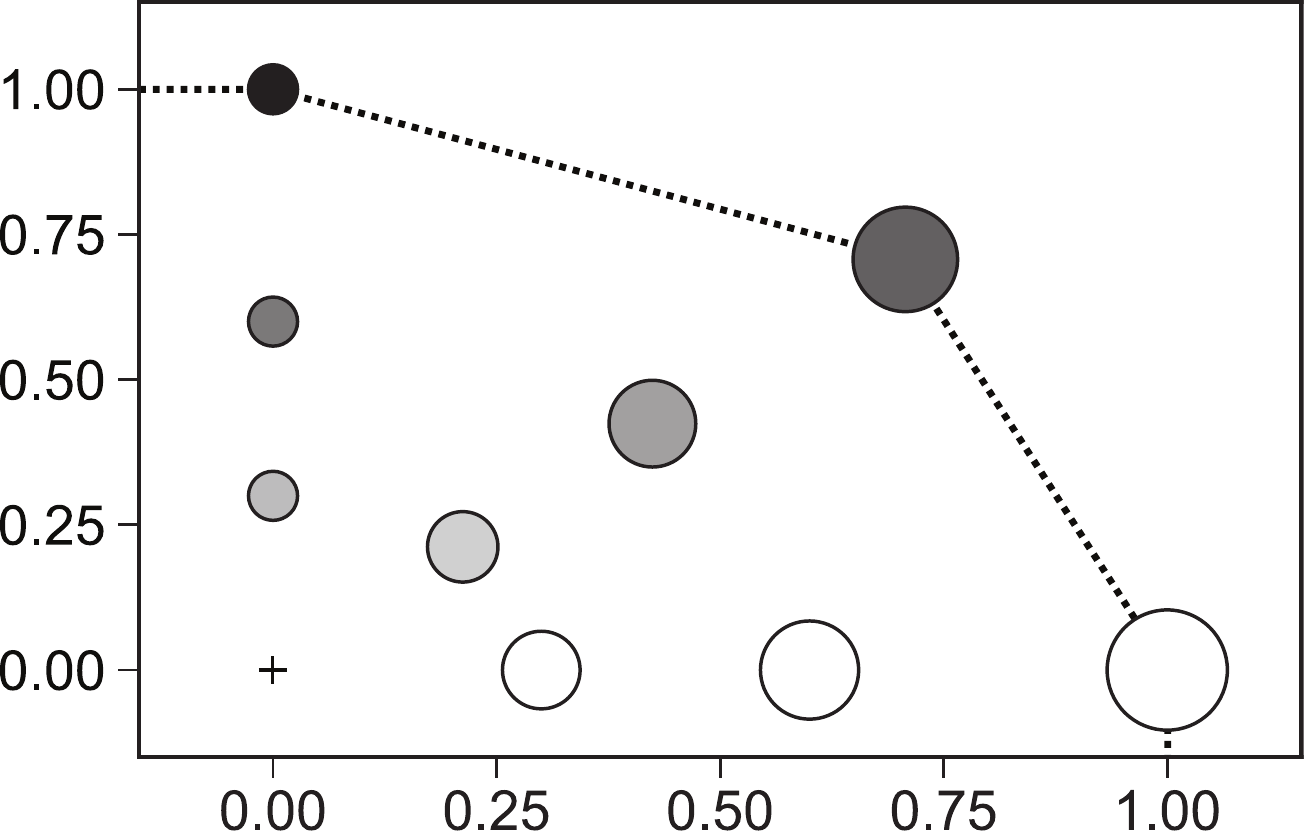}}
    \end{minipage}
    \caption{Illustrative example of the proposed multi-view analysis \textsc{CrowDEA}. (a) Target objects have different sizes and colors. (b) Pairwise comparison is performed by crowd evaluators with individual preferences.
    (c) \textsc{CrowDEA} yields a {\it priority map}, which is a multiple-criteria preference space, where the objects are embedded so that promising candidates are found as the {\it frontier objects}.
    The largest object and the darkest-colored object as well as the fairly large-and-dark object are on the frontier (shown by the dotted line).}\label{fig:overview}
\end{figure*}

We provide an illustrative example in Fig.~\ref{fig:overview}; there are nine objects with different sizes and colors ~(Fig.~\ref{fig:overview}(a)), and we have to prioritize them in terms of various latent criteria, such as size and color.
We ask crowd evaluators to make pairwise preference comparisons based on their own personal criteria~(Fig.~\ref{fig:overview}(b)).
For example, some evaluators prefer darker objects regardless of the object size, whereas others prefer larger objects.
\textsc{CrowDEA} outputs the priority map~(Fig.~\ref{fig:overview}(c)), where the frontier objects are placed on the convex hull (shown by the dotted line) of all the embedded objects.
The $x$-axis is interpreted as the object size and the $y$-axis as the color darkness.
The rightmost and topmost objects are the best according to the size and darkness criteria, respectively. In addition, the top-right object is the best in terms of an intermediate criterion. The object is both fairly large and dark-colored, making it also a promising candidate.

We verify the proposed approach using real datasets that contain numerous ideas or designs. The quantitative results and qualitative analysis demonstrate that \textsc{CrowDEA} outperforms the baselines. The contributions of this study are as follows: 
\begin{itemize}
\item We define a problem that involves organizing and prioritizing a set of ideas from multiple preference viewpoints to support decision-making. 
\item We propose an approach that prioritizes ideas from multiple viewpoints based on pairwise preference comparisons by crowd evaluators. The proposed approach can effectively determine the frontier ideas in a set of ideas. 
\item The embeddings of ideas learned by the proposed approach provide a visualization that facilitates observation of the frontier ideas; in addition, the proposed approach prioritizes ideas from a wider variety of viewpoints, whereas the baselines tend to use the same viewpoints. The proposed approach can also handle various viewpoints and prioritize ideas in situations where only a limited number of evaluators or labels are available. 
\end{itemize}

\section{Related Work}
\subsection{Idea crowdsourcing} 
Existing studies demonstrate that crowdsourcing is an effective method for collecting several creative ideas from a wide range of people ~\cite{yu2011cooks,siangliulue2015toward,prpic2015work}.
To understand a set of ideas, it is important to organize and visualize them. 
Several studies considered with crowdsourcing for organizing ideas. Siangliulue et al. proposed an idea map to visualize a set of ideas using triple-wise similarity queries~\cite{siangliulue2015toward}.
Ahmed and Fuge proposed to find high quality ideas by using community feedback, idea uniqueness, and text features~\cite{10.1145/2998181.2998249}. 
Li et al. proposed an approach that simultaneously ranks and clusters ideas~\cite{SCARPA}. In contrast to these approaches, we allow multiple criteria so that promising candidates can be obtained from various viewpoints (i.e., frontier ideas).
Similar to our work, Lykourentzou et al. proposed a strategy for ranking ideas according to quality and diversity~\cite{10.1145/3274384}.
In their work, the diversity was measured by using the results of manual clustering although our method does not require such manual effort.

\subsection{Decision support methods}
Mathematical methods for supporting decision making have been traditionally studied in operations research. For example, data envelopment analysis (DEA) is a nonparametric method for estimating production frontiers~\cite{DEASurvey,DEABook}, from which the proposed notion of frontier ideas was inspired. The skyline query method, which retains only the objects that are not worse than any others in terms of at least one evaluation criterion, has been extensively studied~\cite{borzsony2001skyline,hose2012survey,crowdskyline}. 
In contrast with DEA and skyline query, the proposed frontier analysis does not require explicit evaluation criteria, and latent evaluation criteria are learned from the data. 

\subsection{Pairwise preference aggregation} 
Methods for aggregating pairwise comparison results have long been discussed.
The Bradley--Terry (BT) model~\cite{bt} is a well-known model for pairwise comparisons.
It estimates a single competency score for each object so that the scores are consistent with the pairwise comparison labels.
To model more complex object relationships, multi-dimensional generalizations of the BT model have been proposed, such as, the multi-dimensional BT model~\cite{2dbt} and intransitivity model~\cite{ChenIntransitivityWSDM,ChenIntransitivityKDD,DuanIntransitivityGeneralized}.
The BT model has also been extended to allow variability in the evaluators~\cite{crowdbt}.
Our work can be considered as the intersection of the above two extensions; we consider multi-dimensional criteria for both evaluators and evaluated objects.

\subsection{Multi-view representation}
In some studies on learning multi-view representations, the term `multi-view' has multiple meanings. 
In several cases, it implies that data instances are described by different types of {explicit} features~\cite{multiviewdata,deepmultiview}, for example, images and texts~\cite{multiviewdata}, texts in two different languages~\cite{multiviewtexttext}, and audio and video media~\cite{multiviewaudiovideo}. 
Amid and Ukkonen targeted multiple {implicit} attributes, where object similarity from triple-wise questions is preserved~\cite{multiview_triplet}.
Their goal is to obtain a space reflecting object similarity, whereas we obtain a space reflecting idea priority.

\subsection{Personalized ranking}
Personalized ranking in recommendation systems in which the relative preference of each user is estimated has been extensively studied. For example, Rendle et al. proposed Bayesian personalized ranking, which trains a matrix factorization model to optimize a ranking loss function~\cite{bpr}. 
This topic has been studied in various scenarios, such as group preference~\cite{gbpr}, visual recommendation~\cite{vbpr}, and event recommendation~\cite{bprnet}.
Their focus is on predicting personalized sets of items for different users, whereas we are interested in obtaining the most advantageous evaluation criterion for each item so that all promising items (i.e., ideas) for decision making may be determined. This results in a different formulation.

\subsection{Search result diversification}
When using web search, users expect not only the most relevant search results to a given query but also diverse ones. Some studies provide both diverse and representative results in terms of content and semantic information~\cite{diverse_representative_www08,diverse_wang2010}, and there are studies on the users' potential intents (such as navigational or informational) of their queries based on their search behaviors~\cite{diverse_intent_www10,diverse_intent_sigir11}. 
An important difference between the abovementioned studies and this one lies in the problem setting: explicit features such as content or context are not available, and we prioritize ideas based solely on pairwise preferences rather than features.
Another difference is that many of these studied have predefined viewpoints, such as the types of user intents, while ours finds the viewpoints from preference comparisons. 

\section{Multi-view Idea Prioritization with Crowds}
\subsection{Models and problem setting}
We address the problem of prioritizing a collection of  $n$ ideas in terms of different latent evaluation criteria. Let $[n] = \{1,2, \cdots, n\}$. Then, we consider the embedding $\bm{x}_i$ for each idea $i \in [n]$ in a $d$-dimensional space, which we call the {\it priority map}.
Each axis of the priority map corresponds to a latent preference criterion, and a large value on an axis implies high preference in terms of the corresponding criterion. 

Decisions are usually made not only according to a single criterion but also by balancing different criteria. For every idea, there should be a viewpoint that best emphasizes its merits, and it is beneficial to determine the set of all ideas that are ``the best'' from certain viewpoints.
We define the best viewpoint for an idea $i$ as a $d$-dimensional unit vector $\bm{v}_i$,
where the projection of $\bm{x}_i$ onto $\bm{v}_i$ (i.e., $\bm{v}_i^\top \bm{x}_i$) is considered to be its preference score from that viewpoint.
If idea $i$ is the most preferred among all the ideas, i.e., $\bm{v}_i^\top\bm{x}_i > \bm{v}_i^\top\bm{x}_j$, for all $j\neq i$, the idea is promising and should be further investigated. The goal is to determine these ideas, which we call \textit{frontier ideas}; they are located on the convex hull (indicated by the dotted line in Fig.~\ref{fig:overview}(c)) of all ideas in the embedding space.
It should be noted that not all ideas can be the best, even from their best viewpoints.

To create the priority map, we collect preference data from $m$ crowd evaluators in the form of pairwise comparisons.
Let $\mathcal{C}_k = \{\left(i, j\right) \mid i, j \in [n], i \succ_k j\}$ be the set of pairwise comparison results by evaluator $k \in [m]$, where $i \succ_k j$ indicates that evaluator $k$ prefers idea $i$ over idea $j$.
As in the case of the best viewpoints for ideas, every crowd evaluator has its individual viewpoint.
We define the viewpoint of crowd evaluator $k$ as a $d$-dimensional unit vector, $\bm{w}_k$.
The projections of $\{\bm{x}_i\}_{i=1}^n$ onto $\bm{w}_k$, i.e., $\{\bm{w}_k^\top \bm{x}_i\}_{i=1}^n$, are regarded as the preference scores by the evaluator, and they are expected to be consistent with the pairwise comparison results, $\mathcal{C}_k$.

In summary, the inputs and outputs of the problem are as follows:
\begin{description}[itemindent=3pt, leftmargin=3pt]
\item[Inputs:] $n$ ideas, $m$ crowd evaluators, and $\{\mathcal{C}_k\}_{k=1}^m$, where 
$\mathcal{C}_k = \{\left(i, j\right) \mid i, j \in [n], i \succ_k j\}$ is the set of pairwise comparison results by evaluator $k\in [m]$.
\item[Outputs:] $\{\bm{x}_i\}_{i=1}^n, \{\bm{v}_i\}_{i=1}^n, \{\bm{w}_k\}_{k=1}^m$, where $\bm{x}_i$ is the $d$-dimensional embedding of idea $i \in [n]$, 
$\bm{v}_i$ is the best viewpoint for idea $i\in[n]$,
and $\bm{w}_k$ is the viewpoint of crowd evaluator $k\in[m]$.
\end{description}

\subsection{Estimation}
We formulate the multi-view analysis as an optimization problem.
Based on the discussions in the previous section,
we have two optimization sub-goals: 
(i) determine as many frontier ideas as possible, and (ii) achieve consistency with the pairwise preference comparison results.

For the first sub-goal, we impose the best viewpoint for each idea, from which the idea is most valuable among all ideas. That is, we require that the resultant idea embeddings $\{\bm{x}_i\}_{i=1}^n$ and corresponding best viewpoints $\{\bm{v}_i\}_{i=1}^n$ satisfy the constraints
\begin{align}
\bm{v}_i^\top\bm{x}_i > \bm{v}_i^\top\bm{x}_j, \forall i \in [n], \forall j\neq i \in[n].\label{eq:lf-constraint}
\end{align}
As it is not possible to satisfy all of the constraints, we quantify the number of constraint violations using a loss function. 
Specifically, we use the hinge loss as the loss function:
\begin{align}
& {\mathcal{L}_{\textrm{F}}}\left(\{\bm{x}_i\}_{i=1}^n, \{\bm{v}_i\}_{i=1}^n\right)= \nonumber \\
& \hspace{10mm} \frac{1}{n(n-1)}\sum_{i \in [n]}\sum_{j \in [n]\setminus i}\max\left\{0,1-\bm{v}_i^\top\left(\bm{x}_i-\bm{x}_j\right)\right\}.
\end{align}

For the second sub-goal, 
the aim is to make the viewpoint of each evaluator consistent with the pairwise comparison results by that evaluator.
We assume that each crowd evaluator has their own viewpoint, and we define ${\bm w}_k$ as the preference criterion vector for the preference labels of evaluator $k$.
From the viewpoint of evaluator $k$, 
the preference score of each idea $i$ is given as $\bm{w}_k^\top \bm{x}_i$;
therefore, the set $\mathcal{C}_k$ of all pairwise comparison results by evaluator $k$ should be consistent with the preference scores, i.e., 
\begin{align}
\bm{w}_k^\top\bm{x}_i > \bm{w}_k^\top\bm{x}_j, 
\forall k \in [m], \forall(i, j) \in \mathcal{C}_k.\label{eq:lc-constraint}
\end{align}

As before, it is not always possible to meet all of the constraints, and again we use the hinge loss function:
\begin{align}
& \mathcal{L}_{\textrm{C}}\left(\{\bm{x}_i\}_{i=1}^n, \{\bm{w}_k\}_{k=1}^m \right)= \nonumber \\
& \hspace{10mm} \frac{1}{c}\sum_{k \in [m]}\sum_{i, j\in\mathcal{C}_k}\max\left\{0,1-\bm{w}_k^\top\left(\bm{x}_i-\bm{x}_j\right)\right\},\label{eq:lc}
\end{align}
where $c = \sum_k |\mathcal{C}_k|$ is the number of observed preference labels.

In addition, we impose the constraint that all embeddings and preference criterion vectors should be non-negative for a more intuitive visualization (as shown in Fig.~\ref{fig:overview}(c)).
Furthermore, we add the constraints that all the preference criterion vectors, $\bm{w}_k$ and $\bm{v}_i$, have unit length, i.e., $\|\bm{w}_k\|_2 = 1$ and $\|\bm{v}_i\|_2 = 1$.
One advantage of this constraint is that it scales the embeddings for all objects. 
This unit length constraint can also avoid the preference criterion vector being zero. For example, for an object $o_i$ that is not on the frontier and ranked low even in its best viewpoint, if $v_i$ is not equal to zero, $\bm{v}_i^\top\left(\bm{x}_i-\bm{x}_j\right)$ for many $o_j$ are lower than zero, which may result in $\bm{v}_i = \bm{0}$ minimizing $\mathcal{L}_{\textrm{F}}\left(\bm{x}_i, \bm{v}_i\right)$.

By combining the loss functions for the two sub-goals and the constraints, 
the optimization problem can be fully formulated as follows:
\begin{align*}
& \minimize_{\{\bm{x}_i\}_{i=1}^n, \{\bm{v}_i\}_{i=1}^n, \{\bm{w}_k\}_{k=1}^m} && \mathcal{L}_{\textrm{C}}\left(\{\bm{x}_i\}_{i=1}^n, \{\bm{w}_k\}_{k=1}^m \right) \\ & && + \alpha {\mathcal{L}_{\textrm{F}}}\left(\{\bm{x}_i\}_{i=1}^n, \{\bm{v}_i\}_{i=1}^n\right) \\
  & \hspace{8mm} \subjectto & & \bm{x}_i, \bm{v}_i, \bm{w}_k \in \mathbb R_+^d, \forall{i \in [n], k \in [m]};\\
  & & & \|\bm{w}_k\|_2 = 1, \forall{k \in [m]};\\
  &&& \|\bm{v}_i\|_2 = 1, \forall{i} \in [n],
\end{align*}
where $\alpha > 0$ is a constant that controls the trade-off between $\mathcal{L}_{\textrm{C}}$ and $\mathcal{L}_{\textrm{F}}$. 

The constrained optimization is performed in a straightforward fashion; after the optimization algorithm updates the parameters at each step, all negative entries are set to zero to satisfy the non-negativity constraints; each $\bm{w}_k$ and $\bm{v}_i$ is then normalized to satisfy the unit length constraint. 
Finally, idea $i$ is considered a frontier idea if 
there exists $\bm{v}$ that satisfies
$\|\bm{v}\|_2 = 1$, $\bm{v} \geq 0$, and $\bm{v}^\top \bm{x}_i > \bm{v}^\top \bm{x}_j$ for all $j\neq i \in [n]$.

\section{Experiments}

\begin{table*}[tb]
\caption{Summary dataset statistics}\label{tbl:dataset}
\vspace{-5mm}
    \begin{subtable}[t]{1\textwidth}
        \centering
        \caption{Ideas}
\begin{tabular}{|l|p{10cm}|l|l|l|}
\hline
Dataset & Problem & \#ideas & \#evaluators & \#labels\\
\hline
\hline
Bike & ``How can we discourage indiscriminate bicycle parking on campus?'' & $81$ & $217$ & $64{,}800$\\
Cheat & ``How can we effectively prevent students from cheating in exams?'' & $80$ & $257$ & $63{,}200$\\
Meeting & ``How can we reduce the number of latecomers for team meetings?'' & $80$ & $177$ & $63{,}200$\\
Night & ``How can we stay safe when walking alone at night?'' & $80$ & $171$ & $63{,}200$\\
Visitor & ``How can we support foreign tourists who encounter a language barrier?'' & $81$ & $158$ & $64{,}800$\\
\hline
\end{tabular}
\end{subtable}
\vfill
    \begin{subtable}[t]{1\textwidth}
        \centering
        \caption{Designs}
\begin{tabular}{|l|p{10cm}|l|l|l|}
\hline
Dataset & Problem & \#ideas & \#evaluators & \#labels\\
\hline
\hline
Olympics & ``Design a logo for the Olympic Games.'' & $38$ & $64$ & $14{,}100$\\
Character & ``Design a character for an AI research laboratory.'' & $66$ & $183$ & $42{,}928$\\
\hline
\end{tabular}
\end{subtable}
\end{table*}

\subsection{Experimental design}
We empirically evaluate the proposed method using real datasets containing ideas and designs for pairwise comparison.
The experiments were designed to answer the following questions:
\begin{enumerate}[label=\textbf{Q\arabic*.}, parsep=0pt] 
\item \textbf{Visualization:} How successful is \textsc{CrowDEA} in organizing ideas?
\item \textbf{Accuracy:} How accurately does \textsc{CrowDEA} prioritize ideas according to multiple viewpoints?
\item \textbf{Efficiency: } How does the accuracy change according to the number of evaluators?
\end{enumerate}

\subsection{Datasets}
We constructed two types of real datasets (Table~\ref{tbl:dataset} summarizes the data statistics)\footnote{Datasets, codes, and Jupyter notebook for reproducing tables and figures are available at: \url{https://github.com/yukinobaba/crowdea}.}:
\begin{itemize}
    \item \textbf{Ideas}: We prepared \if0 eight \fi five open-ended day-to-day life questions, such as ``How can we reduce the number of latecomers at team meetings?'', and we collected solution ideas from crowdsourcing workers using the crowdsourcing platform, Lancers. We obtained approximately $80$ ideas for each question. We hired another set of crowd workers for collecting preference labels, and we asked them to compare pairs of ideas for each problem. Approximately $20$ workers were assigned for each pair of ideas, and each worker evaluated at least $50$ pairs. The order of pairs and that of ideas in each pair were randomized. There were approximately $160$--$260$ evaluators and $64$K preference labels in total for each dataset.
    \item \textbf{Designs: } We held a character design contest for an artificial intelligence (AI) research laboratory and collected $66$ designs. We also prepared $38$ logos for the summer and winter Olympic games from 1948 to 2020, and we collected preference labels for these two design tasks in the same manner as for the datasets containing ideas. 
    There were $183$ evaluators and $43$K preference labels for the ``Character'' dataset and $64$ evaluators and $14$K labels for the ``Olympic" dataset. 
\end{itemize}

\subsection{Baselines}
\begin{table}[tb]
\centering
\caption{Comparison of CrowDEA and baselines}\label{tbl:baseline}
\centering
\begin{tabular}{|c||>{\centering\arraybackslash}p{1.6cm}|>{\centering\arraybackslash}p{1.6cm}|>{\centering\arraybackslash}p{1.2cm}|}
\hline
& Multi- & Multi- & Multi-\\
& evaluators & dimensional & view\\
\hline
\hline
\textsc{BT}\if0~\cite{bt}\fi & - & - & - \\
\textsc{CrowdBT}\if0~\cite{crowdbt}\fi & \checkmark & - & - \\
\textsc{Blade-chest}\if0~\cite{ChenIntransitivityWSDM}\fi & - & \checkmark & - \\
\textsc{BPR}\if0~\cite{bpr}\fi & \checkmark & \checkmark & - \\
\hline
\textsc{CrowDEA} & \checkmark & \checkmark & \checkmark \\
\hline
\end{tabular}
\end{table}

We compare \textsc{CrowDEA} with the following four baselines (They are summarized in Table~\ref{tbl:baseline}): 
\begin{itemize}
\item \textbf{\textsc{BT}}~\cite{bt} is the Bradley--Terry (BT) model, a standard approach for aggregating pairwise preferences. This model represents a preference score for each item by a scalar value and does not assume a different viewpoint for each evaluator. 
\item \textbf{\textsc{CrowdBT}}~\cite{crowdbt} is an extension of \textsc{BT} that incorporates the diversity of evaluator reliability into the model. 
\item \textbf{\textsc{Blade-chest}}~\cite{ChenIntransitivityWSDM} is a multi-dimensional extension of BT and it models intransitivity in pairwise preference.
\item \textbf{\textsc{BPR}}~\cite{bpr} is a method for recommendation, which models both the item embedding and user preference by using $d$-dimensional vectors. 
\end{itemize}

The regularization parameter of the baseline methods was chosen from $\{0.001, 0.01, 0.1\}$, and the best case for a target metric is presented in the results. 
Although there exist several related studies, most of them are not applicable to the present problem setting; only the results of pairwise comparison are given, whereas the features of each idea are unavailable.

\subsection{Setup}
$\alpha$ was set to $0.1$ in all experiments to achieve a good balance between $\mathcal{L}_C$ and $\mathcal{L}_F$.
If $\alpha$ is large, $\mathcal{L}_{\textrm{F}}$ pushes all ideas to the frontier, which does not promote detecting the best ideas, whereas a small $\alpha$ lets the frontier ideas form a small and meaningful subset. As the proposed method aims to generate priority maps, we set $d=2$ or $d=3$.

\subsection{Q1: Visualization}
\begin{figure*}
    \centering
    \begin{subfigure}[t]{0.48\textwidth}
    \begin{center}
        \includegraphics[width=5.5cm]{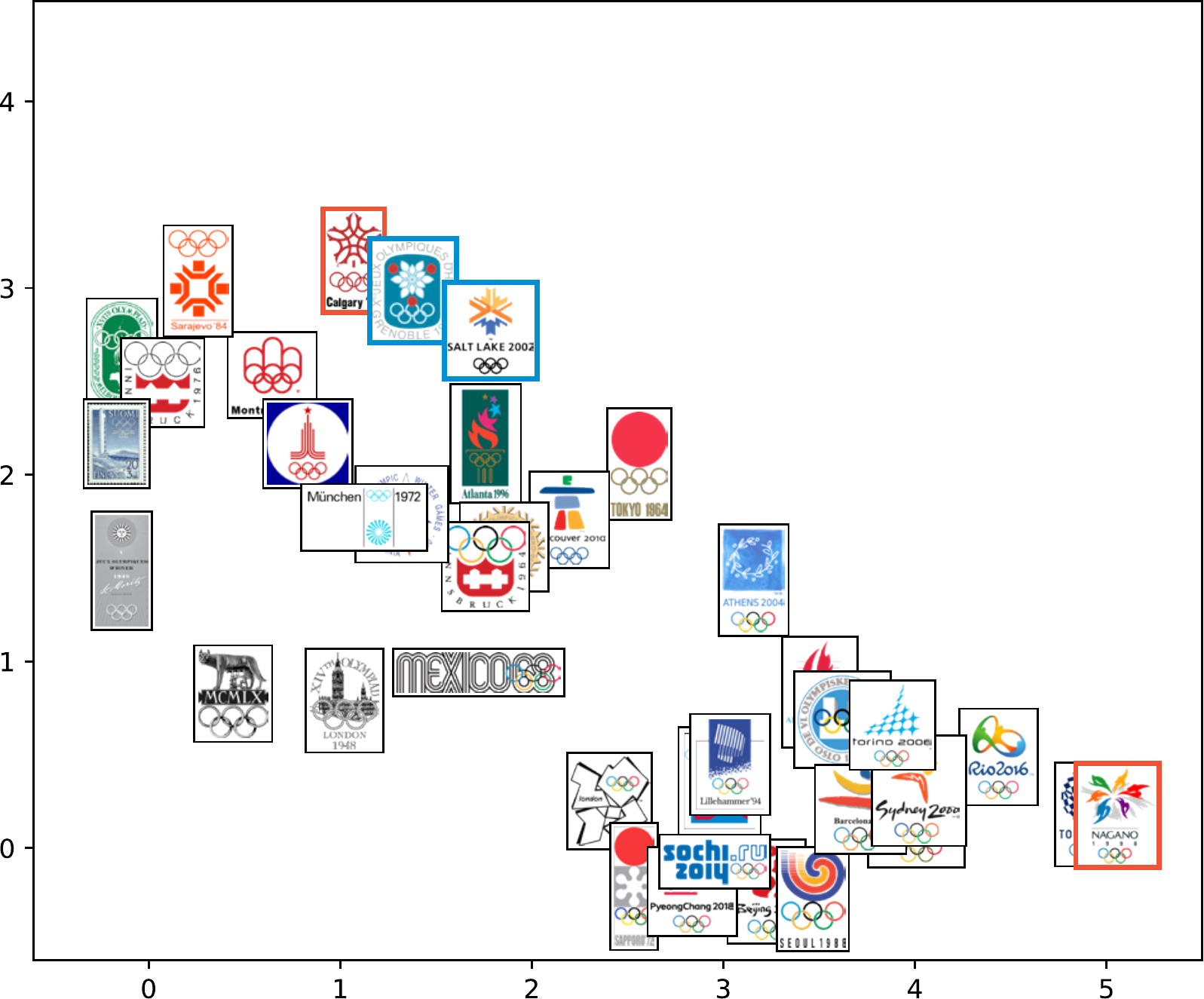}
        \caption{\textsc{CrowDEA}}
        \label{fig:prop-olympic}
    \end{center}
    \end{subfigure}
    \begin{subfigure}[t]{0.48\textwidth}
   	\begin{center}
        \includegraphics[width=5.5cm]{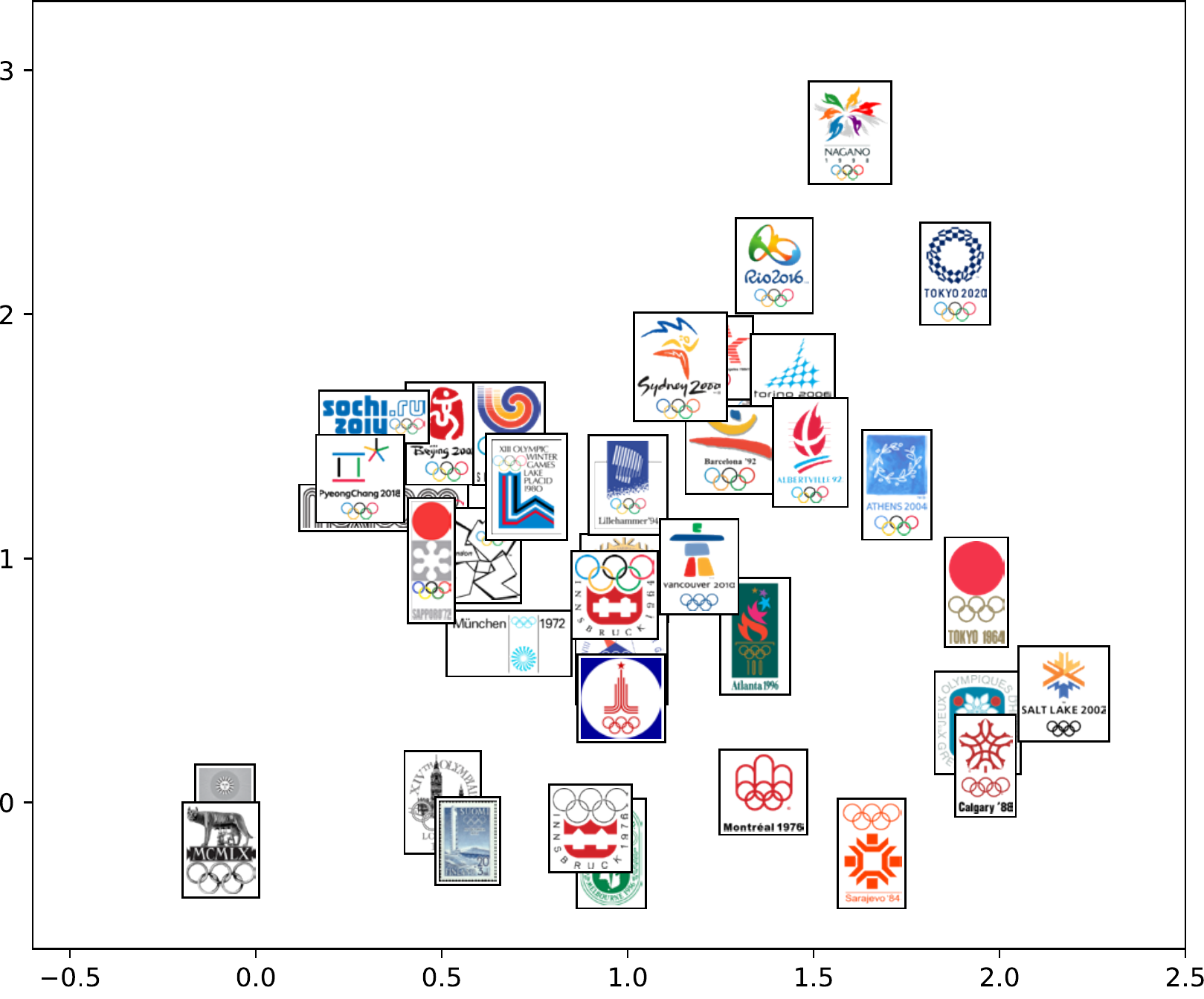}
        \caption{\textsc{BPR}}
        \label{fig:bpr-olympic}
    \end{center}
    \end{subfigure}
    \caption{Priority maps for the ``Olympic'' dataset generated by \textsc{CrowDEA} and \textsc{BPR}. \textsc{CrowDEA} produces well-organized visualization and detects good ideas in diverse viewpoints. 
    The top-right corner of each image corresponds to its embedding in the space. The frontier objects detected by \textsc{CrowDEA} are highlighted in red or blue. Both \textsc{CrowDEA} and \textsc{BPR} locate ideas with higher priorities further from the origin. 
    In contrast to \textsc{BPR}, \textsc{CrowDEA} assigns high priorities to the ideas from multiple viewpoints and organizes the ideas along with the frontier curve.}\label{fig:visualization}
\end{figure*}

We conducted a case study with design datasets to investigate how well \textsc{CrowDEA} visually organizes the ideas from multiple viewpoints. We applied \textsc{CrowDEA} (with $d=2$) to all the preference labels in the dataset, and the estimated two-dimensional embeddings were used for generating the priority map shown in Fig.~\ref{fig:prop-olympic}. 
It can be observed that \textsc{CrowDEA} organizes the ideas along with the frontier curve; 
\textsc{CrowDEA} can locate each idea receiving a higher preference score (from its best viewpoint), and the priority map thus shows the frontier curve. 
This provides a well-organized visualization, which facilitates the evaluation of ideas from multiple viewpoints.
As mentioned in the introduction, the priority map created by \textsc{CrowDEA} allows us to recognize a variety of viewpoints, such as contemporary aesthetics ($x$-axis) and traditional aesthetics ($y$-axis). Recent Olympic logos are placed in the bottom-right region, whereas older logos from the '60s to '80s are placed in the upper-left region, which possibly correlates with the ages of those who provide the preference labels.
It should be noted that the above interpretations of the axes are not given in advance.
In the priority map, Nagano (1998) Olympics\footnote{Nagano (1998) is actually regarded as the best use of athletic imagery by some professional critics, \url{https://en.99designs.jp/blog/famous-design/olympic-logos/}} and Calgary (1988) Olympics, which are highlighted in red, are the two winners on each of the two axes. The priority maps can also capture combinations of these two perspectives, and the winners on them are highlighted in blue in Fig.~\ref{fig:prop-olympic}.
Fig.~\ref{fig:bpr-olympic} shows the visualization produced by \textsc{BPR}, which achieves the highest accuracy, as presented in Sec. 4.6. In contrast to \textsc{CrowDEA}, \textsc{BPR} assigns much higher priorities to modern logos than traditional ones, and it thus does not produce a frontier curve. 

\begin{table}[tb]
\centering
\caption{Examples of frontier ideas for ``Cheat'' problem found by \textsc{CrowDEA}.
\textsc{CrowDEA} finds worthy ideas in various viewpoints.}\label{tbl:frontier}
\centering
\begin{tabular}{|p{7.5cm}|}
\hline
Impose severe penalties for cheating, such as cancellation of modules for an entire year. \\
\hline
Prepare two types of examination sheets with differently ordered items, and distribute one to every student such that neighboring students have different exam sheets. \\
\hline
Have proctors watch students from the back of an examination room. \\
\hline
Instead of multiple-choice questions or short answer questions, use essay questions to make it difficult to copy the answers of other students. \\
\hline
\end{tabular}

\if0
\begin{subtable}[t]{0.45\textwidth}
\centering
\caption{``How can we reduce the number of latecomers for team meetings?'' (Meeting)}
\begin{tabular}{|p{7.5cm}|}
\hline
Schedule a meeting at a time when everyone is supposed to be at the office (for instance, immediately after morning gathering). \\
\hline
Find the right meeting time and cadence for the team.\\
\hline
Impose penalty fees on latecomers.\\
\hline 
Review what to discuss in meetings and make it exciting. \\
\hline
\end{tabular}
\end{subtable}
\fi
\end{table}

\begin{table*}[t]
\footnotesize
\centering
    \caption{Average of nDCG@5 and nDCG@10 scores among the representative viewpoints. \textsc{CrowDEA} accurately ranks the ideas according to various viewpoints. The cases in which \textsc{CrowDEA} outperforms the baselines are bold-faced. The cases in which \textsc{CrowDEA} is the statistically significant ($p < 0.05$) winner by the Wilcoxon signed rank test are underlined.
    }\label{tbl:accuracy}
    \begin{subtable}[t]{1\textwidth}
        \centering
        \caption{$d=2$}
        \label{tbl:accuracy-d2}
\begin{tabular}{|l||c|c|c|c|c|c|c|c|c|c|}
\hline
\multirow{3}{*}{Dataset}&\multicolumn{5}{c|}{nDCG@5}&\multicolumn{5}{c|}{nDCG@10}\\
\cline{2-6}
\cline{7-11}
&\multirow{2}{*}{\textsc{BT}}&\multirow{2}{*}{\textsc{CrowdBT}}&\textsc{Blade}&\multirow{2}{*}{\textsc{BPR}}&\multirow{2}{*}{\textsc{CrowDEA}}&\multirow{2}{*}{\textsc{BT}}&\multirow{2}{*}{\textsc{CrowdBT}}&\textsc{Blade}&\multirow{2}{*}{\textsc{BPR}}&\multirow{2}{*}{\textsc{CrowDEA}}\\
& & &\textsc{-chest}&&&&&\textsc{-chest}&&\\
\hline
\hline
Bike&\underline{${0.772}$}&\underline{${0.779}$}&\underline{${0.757}$}&${0.827}$&$\mathbf{0.833}$&\underline{${0.798}$}&\underline{${0.800}$}&\underline{${0.756}$}&${0.847}$&$\mathbf{0.849}$\\
Cheat&\underline{${0.768}$}&\underline{${0.767}$}&\underline{${0.813}$}&\underline{${0.789}$}&$\mathbf{0.893}$&\underline{${0.795}$}&\underline{${0.791}$}&\underline{${0.819}$}&\underline{${0.800}$}&$\mathbf{0.895}$\\
Meeting&\underline{${0.817}$}&\underline{${0.815}$}&\underline{${0.829}$}&\underline{${0.800}$}&$\mathbf{0.877}$&\underline{${0.824}$}&\underline{${0.825}$}&\underline{${0.837}$}&\underline{${0.818}$}&$\mathbf{0.880}$\\
Night&\underline{${0.790}$}&\underline{${0.790}$}&${0.903}$&\underline{${0.853}$}&$\mathbf{0.917}$&\underline{${0.809}$}&\underline{${0.808}$}&${0.901}$&\underline{${0.862}$}&$\mathbf{0.912}$\\
Visitor&\underline{${0.818}$}&\underline{${0.825}$}&\underline{${0.868}$}&${0.933}$&$\mathbf{0.938}$&\underline{${0.832}$}&\underline{${0.835}$}&\underline{${0.874}$}&${0.938}$&$\mathbf{0.943}$\\
\hline
Character&\underline{${0.902}$}&\underline{${0.912}$}&\underline{${0.866}$}&${0.929}$&$\mathbf{0.930}$&\underline{${0.911}$}&\underline{${0.921}$}&\underline{${0.865}$}&\underline{${0.926}$}&$\mathbf{0.935}$\\
Olympic&${0.926}$&${0.926}$&${0.920}$&$\mathbf{0.940}$&${0.936}$&${0.937}$&${0.937}$&${0.923}$&$\mathbf{0.949}$&${0.947}$\\
\hline
\end{tabular}
\end{subtable}
\vfill
\begin{subtable}[t]{1\textwidth}
\centering
\caption{$d=3$}
\label{tbl:accuracy-d3}
\begin{tabular}{|l||c|c|c|c|c|c|c|c|c|c|}
\hline
\multirow{3}{*}{Dataset}&\multicolumn{5}{c|}{nDCG@5}&\multicolumn{5}{c|}{nDCG@10}\\
\cline{2-6}
\cline{7-11}
&\multirow{2}{*}{\textsc{BT}}&\multirow{2}{*}{\textsc{CrowdBT}}&\textsc{Blade}&\multirow{2}{*}{\textsc{BPR}}&\multirow{2}{*}{\textsc{CrowDEA}}&\multirow{2}{*}{\textsc{BT}}&\multirow{2}{*}{\textsc{CrowdBT}}&\textsc{Blade}&\multirow{2}{*}{\textsc{BPR}}&\multirow{2}{*}{\textsc{CrowDEA}}\\
& & &\textsc{-chest}&&&&&\textsc{-chest}&&\\
\hline
\hline
Bike&\underline{${0.772}$}&\underline{${0.779}$}&\underline{${0.803}$}&\underline{${0.819}$}&$\mathbf{0.883}$&\underline{${0.798}$}&\underline{${0.800}$}&\underline{${0.797}$}&\underline{${0.835}$}&$\mathbf{0.893}$\\
Cheat&\underline{${0.768}$}&\underline{${0.767}$}&\underline{${0.847}$}&\underline{${0.795}$}&$\mathbf{0.924}$&\underline{${0.795}$}&\underline{${0.791}$}&\underline{${0.839}$}&\underline{${0.804}$}&$\mathbf{0.927}$\\
Meeting&\underline{${0.817}$}&\underline{${0.815}$}&\underline{${0.867}$}&\underline{${0.888}$}&$\mathbf{0.920}$&\underline{${0.824}$}&\underline{${0.825}$}&\underline{${0.862}$}&\underline{${0.891}$}&$\mathbf{0.923}$\\
Night&\underline{${0.790}$}&\underline{${0.790}$}&\underline{${0.907}$}&\underline{${0.913}$}&$\mathbf{0.953}$&\underline{${0.809}$}&\underline{${0.808}$}&\underline{${0.894}$}&\underline{${0.910}$}&$\mathbf{0.945}$\\
Visitor&\underline{${0.818}$}&\underline{${0.825}$}&\underline{${0.916}$}&\underline{${0.842}$}&$\mathbf{0.955}$&\underline{${0.832}$}&\underline{${0.835}$}&\underline{${0.906}$}&\underline{${0.846}$}&$\mathbf{0.951}$\\
\hline
Character&\underline{${0.902}$}&\underline{${0.912}$}&\underline{${0.905}$}&${0.957}$&$\mathbf{0.960}$&${0.911}$&${0.921}$&${0.891}$&$\mathbf{0.954}$&${0.953}$\\
Olympic&\underline{${0.926}$}&\underline{${0.926}$}&\underline{${0.927}$}&\underline{${0.956}$}&$\mathbf{0.966}$&\underline{${0.937}$}&\underline{${0.937}$}&\underline{${0.926}$}&\underline{${0.952}$}&$\mathbf{0.964}$\\
\hline
\end{tabular}
\end{subtable}
\end{table*}

\subsection{Q2: Accuracy}
We demonstrate how accurately \textsc{CrowDEA} determines the best ideas in various viewpoints.

\textbf{Setup:} We prepared the ground truth of idea priorities from various viewpoints to investigate accuracy. We first collected $100$ viewpoints for each dataset from crowdsourcing workers who were shown a pair of ideas and asked to describe a viewpoint that distinguishes the two ideas. For instance, we obtained ``This idea can be easily implemented'' as a viewpoint for the ``Cheat'' problem. We then asked workers to grade each idea in terms of each viewpoint on a five-point scale. Ten workers were assigned to each idea--viewpoint pair, and the average grade was used as the ground truth priority $p^*_{ij}$ of idea $i$ from viewpoint $j$. We removed overlapped or less popular viewpoints by applying $k$-means clustering to the obtained priorities; that is, we considered $\bm{p}^*_j = \left(p^*_{1j}, \ldots, p^*_{nj}\right)$ to be the feature vector of viewpoint $j$ and used it for clustering. The number of clusters was set to $50$\footnote{The representative viewpoints were almost the same when the number of clusters was chosen from $\{30, 40, 50, 60, 70\}$.}. The clusters with only one sample were then omitted, and the number of remaining clusters was $15$--$30$. We chose the viewpoint closest to the center of each of the remaining clusters, referred to as a representative viewpoint. We thus had $15$--$30$ representative viewpoints for each dataset. We note that neither the proposed method nor the baseline methods can access the ground truth; it is used only for evaluation. 

We applied \textsc{CrowDEA}, \textsc{BPR}, and \textsc{Blade-chest} to the preference labels in each dataset and obtained the embeddings $\{\bm{x}_i\}_{i=1}^n$. We intended to use the embeddings to rank ideas according to each representative viewpoint in the ground truth to evaluate the ranking accuracy. Given a viewpoint vector $\bm{v}$, the projection of $\bm{x}_i$ onto $\bm{v}$ (i.e., $\bm{v}^\top \bm{x}_i$) is considered as the priority score for this viewpoint. We optimize a viewpoint vector $\bm{v}^*_j$, which well represents viewpoint $j$, according to a evaluation measure. 
This yielded $p_{ij} = \bm{v}_j^{*\top}\bm{x}_i$, which is the predicted priority score for that viewpoint. 
We also applied \textsc{BT} and \textsc{CrowdBT} to the preference labels, and regarded the estimated score $p_i$ as $p_{ij}$ for each viewpoint $j$. 
Each method generated a ranking of the ideas for viewpoint $j$ according to $\{p_{ij}\}_i$. We compared these with the ranking by the ground truth priorities, $\{p^*_{ij}\}_i$, and evaluated the ranking accuracy.

The ranking accuracy (nDCG@$k$) for each viewpoint was calculated as follows: we had the top $k$ ideas according to the predicted priorities, and their true priorities, $\bm{y}=(y_1, \ldots, y_k)$, where $y_i$ is the true priority of the $i$-th ranked idea. We additionally had the true top $k$ ideas and their true priorities $\bm{t}=(t_1, \ldots, t_k)$. We calculated $\mathrm{DCG}(k,\bm{y}) = \sum^k_{i=1} y_{i} / \log_2(i+1)$ and $\mathrm{IDCG}(k,\bm{t}) = \sum^k_{i=1} t_{i} / \log_2(i+1)$ to obtain $\mathrm{nDCG@}k = \mathrm{DCG}(k,\bm{y}) / \mathrm{IDCG}(k,\bm{t})$.

\begin{figure*}
    \centering
    \vspace{5mm}
    \begin{subfigure}[t]{1\textwidth}
    \begin{center}
        \includegraphics[width=16cm]{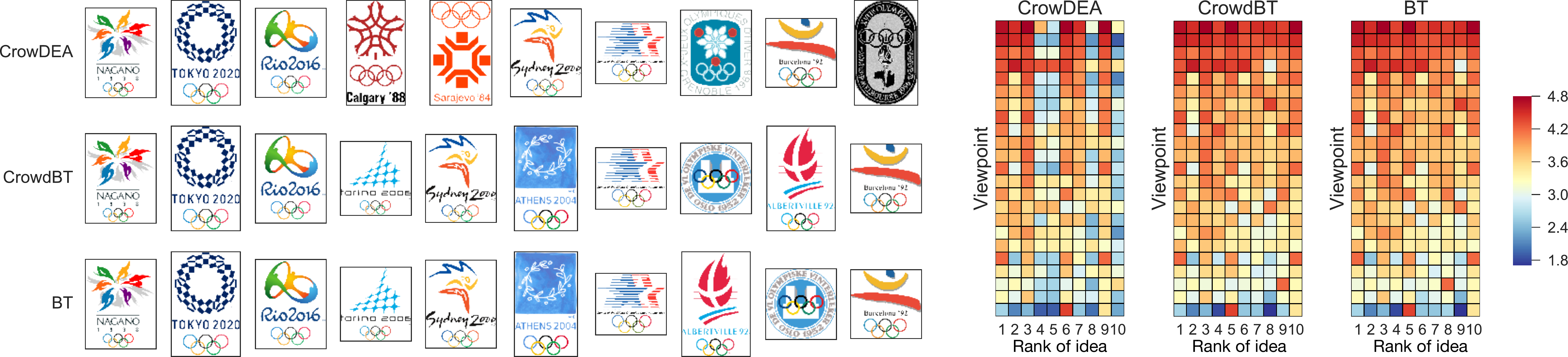}
        \caption{``Olympics''}
        \label{fig:ranking-olympic}
    \end{center}
    \end{subfigure}
    \vfill
    \vspace{5mm}
    \begin{subfigure}[t]{1\textwidth}
    \begin{center}
        \includegraphics[width=16cm]{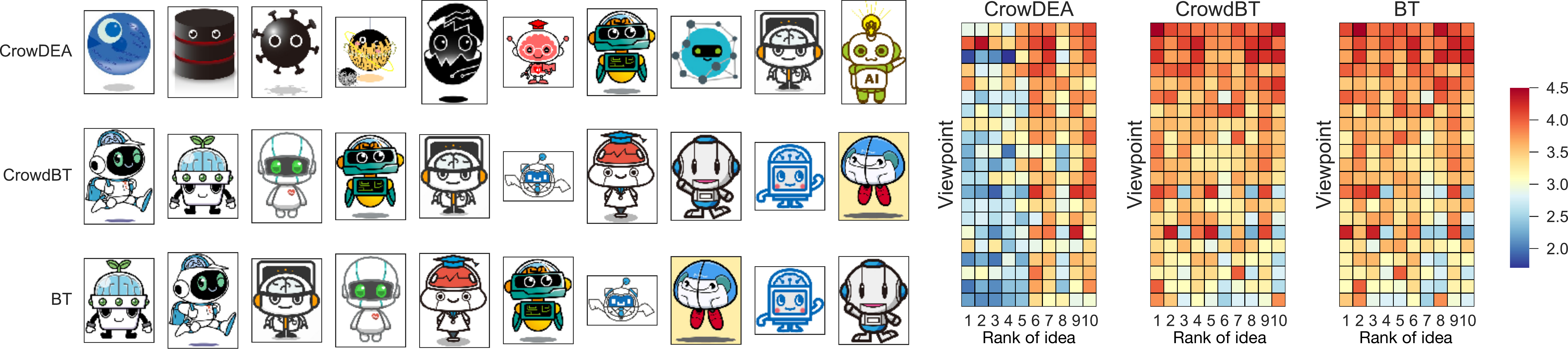}
        \caption{``Character''}
        \label{fig:ranking-character}
    \end{center}
    \end{subfigure}
    \caption{
    (Left) The top-10 ideas prioritized by each method. The ideas are ordered from left to right according to their estimated preference scores. (Right) The ground truth priority of each of the top-10 ideas in each representative viewpoint. The ideas selected by \textsc{CrowDEA} are prioritized in different viewpoints, while those chosen by the baselines are prioritized in the same viewpoints.}\label{fig:ranking}
\end{figure*}

\textbf{Results:} Table~\ref{tbl:frontier} lists examples of the frontier ideas obtained by \textsc{CrowDEA} for the ``Cheat'' dataset. It can be seen that \textsc{CrowDEA} provides useful ideas that are considered good from various viewpoints. Table~\ref{tbl:accuracy} shows the average nDCG@5 and nDCG@10 over the representative viewpoints. It can be seen that \textsc{CrowDEA} outperforms the baselines in most cases; \textsc{CrowDEA} can capture the diversity of viewpoints that are not considered by the other simple methods. Moreover, \textsc{CrowDEA} with $d=3$ achieves higher scores than with $d=2$ in all datasets, as the higher-dimensional embedding handles various viewpoints. 

We quantitatively investigate the variety of the ideas prioritized by the proposed method. Fig.~\ref{fig:ranking} shows the top-$10$ ideas and a heatmap of the ground truth priority $p_{ij}^*$ of each top-$10$ idea for each viewpoint. The top-$10$ ideas ranked by \textsc{CrowdBT} (with $\lambda=0.01$) and \textsc{BT} ($\lambda=0.01$) are selected by using $p_i$, and those by \textsc{CrowDEA} (with $d=2$) are according to $p_i = \sum_{j\in [n]\setminus i}\bm{v}_i^\top(\bm{x}_i - \bm{x}_j)$, which indicates how likely the ideas are to be frontier ideas. It is observed that \textsc{CrowDEA} prioritizes ideas from a wider variety of viewpoints, whereas the baselines tend to use the same viewpoints. Note that \textsc{Blade-chest} and \textsc{BPR} cannot output a single priority score due to the absence of $\bm{v}_i$. 

\subsection{Q3: Efficiency}
Each dataset contains the preference labels from approximately 200 evaluators; however, it is not always feasible to collect these labels from a large group of evaluators. To demonstrate the efficiency of the proposed method, we evaluate the accuracy of \textsc{CrowDEA} in terms of the number of evaluators. Additionally, each dataset contains $200$--$400$ labels per evaluator. We evaluate the accuracy of \textsc{CrowDEA} in cases where the number of available labels is limited.

\textbf{Setup:} We randomly chose $q \in \{20, 50, 100\}$ evaluators or $r \in \{1000, 2000, 5000, 10000, 20000\}$ labels and applied \textsc{CrowDEA} to the preference labels (i.e., a subset of the preference labels in a dataset). For each $q$ or $r$, we performed $10$ trials and selected a different set of evaluators (or labels) for each trial. 

\begin{figure*}
    \centering
    \begin{subfigure}[t]{1\textwidth}
        \centering
        \includegraphics[width=15cm]{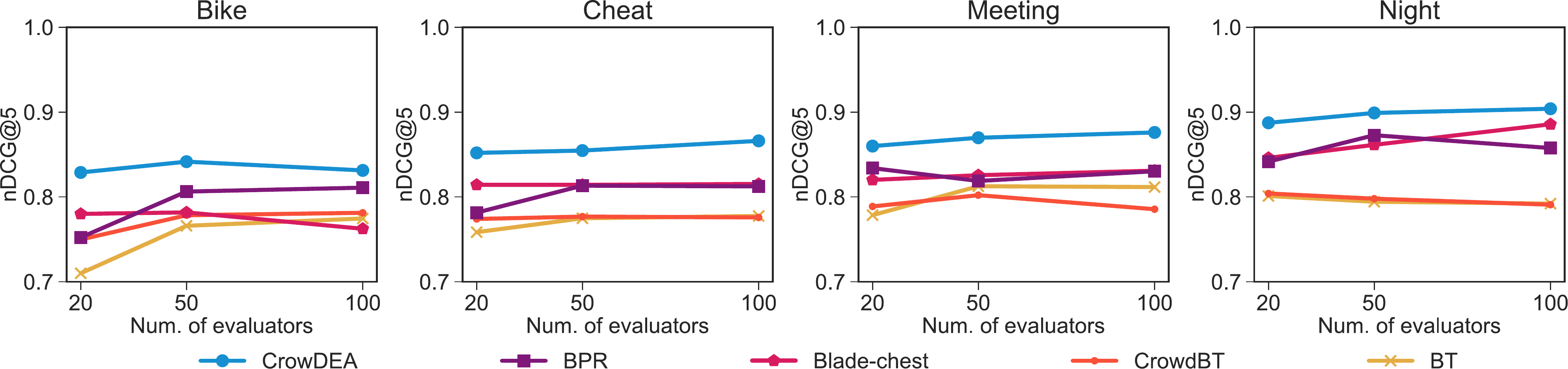}
        \caption{Efficiency with a small number of evaluators}\label{fig:efficiency-evaluator}
        \vspace{10mm}
    \end{subfigure}
    \begin{subfigure}[t]{1\textwidth}
        \centering
        \includegraphics[width=16cm]{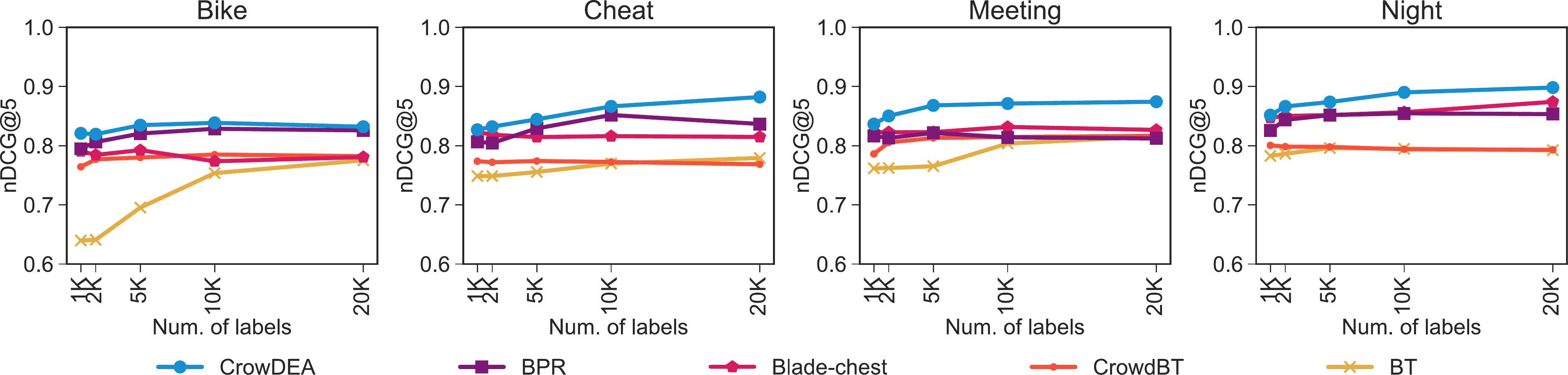}
        \caption{Efficiency with a small number of labels}\label{fig:efficiency-label}
    \end{subfigure}
    \caption{Average of nDCG@5 scores for the representative viewpoints and ten trials. \textsc{CrowDEA} accurately ranks the ideas even when the number of evaluators or the number of labels is small. $d$ is set to $2$. Due to space limitation, we only present the results of the first four datasets. 
    }\label{fig:efficiency}
\end{figure*}

\textbf{Results:} Fig.~\ref{fig:efficiency-evaluator} shows the average nDCG@5 of each method according to the number of evaluators used for model inference. The average nDCG@5 scores are shown for different viewpoints and ten different subsets of evaluators. The performance of \textsc{CrowDEA} declines as the number of evaluators decreases; however, the average nDCG@5 scores are still over $0.8$ in all cases, even when the number of evaluators is only $20$, and \textsc{CrowDEA} outperforms the baselines in all cases. Fig.~\ref{fig:efficiency-label} shows the average nDCG@5 of each method according to the number of labels. \textsc{CrowDEA} shows better performance than the other methods even when the number of labels is small. It is worth noting that \textsc{CrowDEA} can handle various viewpoints and prioritize ideas in situations where only a limited number of evaluators or labels are available.

\section{Conclusions}
We addressed the problem of idea prioritization with crowds. The proposed method estimates the best viewpoint for every idea and preference criterion of every crowd evaluator. Experimental results based on real datasets containing ideas demonstrated that the proposed approach effectively prioritizes ideas from multiple viewpoints and obtains frontier ideas. The visualization based on the learned embeddings facilitates observation of the frontier ideas. Possible future work may include extensions to multiple best viewpoints for each idea, as the present formulation allows only a single best viewpoint.
The interpretation of the obtained results is also an important issue; although this is left to users in the present study, systematic interpretation by crowds is an interesting future research direction.

\section*{Acknowledgments}
This work was supported by JSPS KAKENHI Grant Number JP18K18105 and JST PRESTO Grant Number JPMJPR19J9, Japan.

\clearpage
\bibliographystyle{aaai}
\bibliography{reference}

\begin{thebibliography}{}

\bibitem[\protect\citeauthoryear{Ahmed and
  Fuge}{2017}]{10.1145/2998181.2998249}
Ahmed, F., and Fuge, M.
\newblock 2017.
\newblock Capturing winning ideas in online design communities.
\newblock In {\em Proceedings of the 2017 ACM Conference on Computer Supported
  Cooperative Work and Social Computing (CSCW)},  1675--1687.

\bibitem[\protect\citeauthoryear{Amid and Ukkonen}{2015}]{multiview_triplet}
Amid, E., and Ukkonen, A.
\newblock 2015.
\newblock Multiview triplet embedding: learning attributes in multiple maps.
\newblock In {\em Proceedings of the 32nd International Conference on Machine
  Learning (ICML)},  1472--1480.

\bibitem[\protect\citeauthoryear{Borzsony, Kossmann, and
  Stocker}{2001}]{borzsony2001skyline}
Borzsony, S.; Kossmann, D.; and Stocker, K.
\newblock 2001.
\newblock The skyline operator.
\newblock In {\em Proceedings of the 17th International Conference on Data
  Engineering (ICDE)},  421--430.

\bibitem[\protect\citeauthoryear{Bradley and Terry}{1952}]{bt}
Bradley, R.~A., and Terry, M.~E.
\newblock 1952.
\newblock Rank analysis of incomplete block designs: I. the method of paired
  comparisons.
\newblock {\em Biometrika} 39(3/4):324--345.

\bibitem[\protect\citeauthoryear{Causeur and Husson}{2005}]{2dbt}
Causeur, D., and Husson, F.
\newblock 2005.
\newblock A 2-dimensional extension of the bradley--terry model for paired
  comparisons.
\newblock {\em Journal of Statistical Planning and Inference} 135(2):245--259.

\bibitem[\protect\citeauthoryear{Chandar \bgroup et al\mbox.\egroup
  }{2014}]{multiviewtexttext}
Chandar, S.; Lauly, S.; Larochelle, H.; Khapra, M.; Ravindran, B.; Raykar,
  V.~C.; and Saha, A.
\newblock 2014.
\newblock An autoencoder approach to learning bilingual word representations.
\newblock In {\em Advances in Neural Information Processing Systems 27},
  1853--1861.

\bibitem[\protect\citeauthoryear{Chen and
  Joachims}{2016a}]{ChenIntransitivityWSDM}
Chen, S., and Joachims, T.
\newblock 2016a.
\newblock Modeling intransitivity in matchup and comparison data.
\newblock In {\em Proceedings of the 9th ACM International Conference on Web
  Search and Data Mining (WSDM)},  227--236.

\bibitem[\protect\citeauthoryear{Chen and
  Joachims}{2016b}]{ChenIntransitivityKDD}
Chen, S., and Joachims, T.
\newblock 2016b.
\newblock Predicting matchups and preferences in context.
\newblock In {\em Proceedings of the 22nd ACM SIGKDD International Conference
  on Knowledge Discovery and Data Mining (KDD)},  775--784.

\bibitem[\protect\citeauthoryear{Chen \bgroup et al\mbox.\egroup
  }{2013}]{crowdbt}
Chen, X.; Bennett, P.~N.; Collins-Thompson, K.; and Horvitz, E.
\newblock 2013.
\newblock Pairwise ranking aggregation in a crowdsourced setting.
\newblock In {\em Proceedings of the 6th ACM International Conference on Web
  Search and Data Mining (WSDM)},  193--202.

\bibitem[\protect\citeauthoryear{Cheng, Gao, and
  Liu}{2010}]{diverse_intent_www10}
Cheng, Z.; Gao, B.; and Liu, T.-Y.
\newblock 2010.
\newblock Actively predicting diverse search intent from user browsing
  behaviors.
\newblock In {\em Proceedings of the 19th International Conference on World
  Wide Web (WWW)},  221--230.

\bibitem[\protect\citeauthoryear{Cooper, Seiford, and Zhu}{2004}]{DEABook}
Cooper, W.~W.; Seiford, L.~M.; and Zhu, J.
\newblock 2004.
\newblock Data envelopment analysis.
\newblock In {\em Handbook on Data Envelopment Analysis}. Springer.
\newblock  1--39.

\bibitem[\protect\citeauthoryear{Duan \bgroup et al\mbox.\egroup
  }{2017}]{DuanIntransitivityGeneralized}
Duan, J.; Li, J.; Baba, Y.; and Kashima, H.
\newblock 2017.
\newblock A generalized model for multidimensional intransitivity.
\newblock In {\em In Proceedings of the 21st Pacific-Asia Conference on
  Knowledge Discovery and Data Mining (PAKDD)},  840--852.

\bibitem[\protect\citeauthoryear{He and McAuley}{2016}]{vbpr}
He, R., and McAuley, J.
\newblock 2016.
\newblock {VBPR}: Visual {B}ayesian personalized ranking from implicit
  feedback.
\newblock In {\em Proceedings of the 30th AAAI Conference on Artificial
  Intelligence (AAAI)},  144--150.

\bibitem[\protect\citeauthoryear{Hose and Vlachou}{2012}]{hose2012survey}
Hose, K., and Vlachou, A.
\newblock 2012.
\newblock A survey of skyline processing in highly distributed environments.
\newblock {\em The VLDB Journal} 21(3):359--384.

\bibitem[\protect\citeauthoryear{Huang and
  Kingsbury}{2013}]{multiviewaudiovideo}
Huang, J., and Kingsbury, B.
\newblock 2013.
\newblock Audio-visual deep learning for noise robust speech recognition.
\newblock In {\em Proceedings of 2013 IEEE International Conference on
  Acoustics, Speech and Signal Processing (ICASSP)},  7596--7599.

\bibitem[\protect\citeauthoryear{Kennedy and
  Naaman}{2008}]{diverse_representative_www08}
Kennedy, L.~S., and Naaman, M.
\newblock 2008.
\newblock Generating diverse and representative image search results for
  landmarks.
\newblock In {\em Proceedings of the 17th International Conference on World
  Wide Web (WWW)},  297--306.

\bibitem[\protect\citeauthoryear{Koyama, Sakamoto, and
  Igarashi}{2014}]{koyama2014crowd}
Koyama, Y.; Sakamoto, D.; and Igarashi, T.
\newblock 2014.
\newblock Crowd-powered parameter analysis for visual design exploration.
\newblock In {\em Proceedings of the 27th Annual ACM Symposium on User
  Interface Software and Technology (UIST)},  65--74.

\bibitem[\protect\citeauthoryear{Li, Baba, and Kashima}{2018}]{SCARPA}
Li, J.; Baba, Y.; and Kashima, H.
\newblock 2018.
\newblock Simultaneous clustering and ranking from pairwise comparisons.
\newblock In {\em Proceedings of the 27th International Joint Conference on
  Artificial Intelligence (IJCAI)},  1554--1560.

\bibitem[\protect\citeauthoryear{Li, Yang, and Zhang}{2016}]{multiviewdata}
Li, Y.; Yang, M.; and Zhang, Z.
\newblock 2016.
\newblock Multi-view representation learning: A survey from shallow methods to
  deep methods.
\newblock {\em arXiv preprint arXiv:1610.01206}.

\bibitem[\protect\citeauthoryear{Lofi, El~Maarry, and
  Balke}{2013}]{crowdskyline}
Lofi, C.; El~Maarry, K.; and Balke, W.-T.
\newblock 2013.
\newblock Skyline queries in crowd-enabled databases.
\newblock In {\em Proceedings of the 16th International Conference on Extending
  Database Technology (EDBT)},  465--476.

\bibitem[\protect\citeauthoryear{Lykourentzou \bgroup et al\mbox.\egroup
  }{2018}]{10.1145/3274384}
Lykourentzou, I.; Ahmed, F.; Papastathis, C.; Sadien, I.; and Papangelis, K.
\newblock 2018.
\newblock When crowds give you lemons: Filtering innovative ideas using a
  diverse-bag-of-lemons strategy.
\newblock {\em Proceedings of the ACM Human Computer Interaction}.

\bibitem[\protect\citeauthoryear{Pan and Chen}{2013}]{gbpr}
Pan, W., and Chen, L.
\newblock 2013.
\newblock {GBPR}: Group preference based bayesian personalized ranking for
  one-class collaborative filtering.
\newblock In {\em Proceedings of the 23rd International Joint Conference on
  Artificial Intelligence (IJCAI)},  2691--2697.

\bibitem[\protect\citeauthoryear{Prpi{\'c} \bgroup et al\mbox.\egroup
  }{2015}]{prpic2015work}
Prpi{\'c}, J.; Shukla, P.~P.; Kietzmann, J.~H.; and McCarthy, I.~P.
\newblock 2015.
\newblock How to work a crowd: Developing crowd capital through crowdsourcing.
\newblock {\em Business Horizons} 58(1):77--85.

\bibitem[\protect\citeauthoryear{Qiao \bgroup et al\mbox.\egroup
  }{2014}]{bprnet}
Qiao, Z.; Zhang, P.; Zhou, C.; Cao, Y.; Guo, L.; and Zhang, Y.
\newblock 2014.
\newblock Event recommendation in event-based social networks.
\newblock In {\em Proceedings of the 28th AAAI Conference on Artificial
  Intelligence (AAAI)},  3130--3131.

\bibitem[\protect\citeauthoryear{Rendle \bgroup et al\mbox.\egroup
  }{2009}]{bpr}
Rendle, S.; Freudenthaler, C.; Gantner, Z.; and Schmidt-Thieme, L.
\newblock 2009.
\newblock {BPR}: {B}ayesian personalized ranking from implicit feedback.
\newblock In {\em Proceedings of the 25th Conference on Uncertainty in
  Artificial Intelligence (UAI)},  452--461.

\bibitem[\protect\citeauthoryear{Santos, Macdonald, and
  Ounis}{2011}]{diverse_intent_sigir11}
Santos, R.~L.; Macdonald, C.; and Ounis, I.
\newblock 2011.
\newblock Intent-aware search result diversification.
\newblock In {\em Proceedings of the 34th International ACM SIGIR Conference on
  Research and Development in Information Retrieval (SIGIR)},  595--604.

\bibitem[\protect\citeauthoryear{Seiford and Thrall}{1990}]{DEASurvey}
Seiford, L.~M., and Thrall, R.~M.
\newblock 1990.
\newblock Recent developments in dea: the mathematical programming approach to
  frontier analysis.
\newblock {\em Journal of Econometrics} 46(1-2):7--38.

\bibitem[\protect\citeauthoryear{Siangliulue \bgroup et al\mbox.\egroup
  }{2015}]{siangliulue2015toward}
Siangliulue, P.; Arnold, K.~C.; Gajos, K.~Z.; and Dow, S.~P.
\newblock 2015.
\newblock Toward collaborative ideation at scale: Leveraging ideas from others
  to generate more creative and diverse ideas.
\newblock In {\em Proceedings of the 18th ACM Conference on Computer Supported
  Cooperative Work and Social Computing (CSCW)},  937--945.

\bibitem[\protect\citeauthoryear{Wang \bgroup et al\mbox.\egroup
  }{2010}]{diverse_wang2010}
Wang, M.; Yang, K.; Hua, X.-S.; and Zhang, H.-J.
\newblock 2010.
\newblock Towards a relevant and diverse search of social images.
\newblock {\em IEEE Transactions on Multimedia} 12(8):829--842.

\bibitem[\protect\citeauthoryear{Wang \bgroup et al\mbox.\egroup
  }{2015}]{deepmultiview}
Wang, W.; Arora, R.; Livescu, K.; and Bilmes, J.
\newblock 2015.
\newblock On deep multi-view representation learning.
\newblock In {\em Proceedings of the 32nd International Conference on
  International Conference on Machine Learning (ICML)},  1083--1092.

\bibitem[\protect\citeauthoryear{Yu and Nickerson}{2011}]{yu2011cooks}
Yu, L., and Nickerson, J.~V.
\newblock 2011.
\newblock Cooks or cobblers?: crowd creativity through combination.
\newblock In {\em Proceedings of the SIGCHI Conference on Human Factors in
  Computing Systems (CHI)},  1393--1402.

\end{thebibliography}

\end{document}